\documentclass[twocolumn,english,showpacs,nofootinbib,notitlepage,superscriptaddress]{revtex4-2}
\usepackage{color}
\usepackage{graphicx}
\usepackage{bbold}
\usepackage{bm}
\usepackage{amsmath}
\usepackage{amsfonts}
\usepackage{amssymb}
\usepackage{braket}
\usepackage{units}
\usepackage{csquotes}
\usepackage{upgreek}
\usepackage{xcolor}
\usepackage{fancyhdr}
\usepackage{float}
\usepackage{lipsum}
\usepackage{printlen}
\begin{document}
	
%	\pagestyle{fancy}
%	\chead{draft \today}
%	\rhead{\thepage}
%	\renewcommand{\headrulewidth}{0.0pt} 

\title{Cooling a strongly-interacting quantum gas by  interaction modulation}

\author{D.~Eberz}
\affiliation{Physikalisches Institut, University of Bonn, Wegelerstra{\ss}e 8, 53115 Bonn, Germany}
\author{A. Kell}
\affiliation{Physikalisches Institut, University of Bonn, Wegelerstra{\ss}e 8, 53115 Bonn, Germany}
\author{M. Breyer}
\affiliation{Physikalisches Institut, University of Bonn, Wegelerstra{\ss}e 8, 53115 Bonn, Germany}
\author{M. Köhl}
\affiliation{Physikalisches Institut, University of Bonn, Wegelerstra{\ss}e 8, 53115 Bonn, Germany}

\begin{abstract}
{ We present a cooling method for a strongly-interacting trapped quantum gas. By applying a magnetic field modulation with frequencies close to the binding energy of a molecular bound state we selectively remove dimers with high kinetic energy from the sample. We demonstrate cooling of the sample over a wide range of interaction strengths and measure a high cooling efficiency of $\gamma=4$ that exceeds all previous cooling near Feshbach resonances.}\end{abstract}
\maketitle

Experiments with trapped ultracold quantum gases have been made possible by groundbreaking advancements in cooling techniques. % \cite{ketterle_evaporative_1996,chu_nobel_1998,cohen-tannoudji_nobel_1998,phillips_nobel_1998}. 
This has led, for example, to the observation of Bose-Einstein condensation (BEC) \cite{anderson_observation_1995}, degenerate Fermi gases \cite{demarco_onset_1999} and strongly-correlated quantum states such as Mott insulators \cite{greiner_quantum_2002} and the unitary Fermi gas \cite{regal_observation_2004,kinast_evidence_2004}. Theoretically, many more complex quantum states are expected at even lower temperature, for example, unconventional superconductivity \cite{Sigrist1991}, topological quantum states {\cite{Kitaev2001} or spin liquids \cite{Knolle2019}. However, as we approach very low temperatures, efficient cooling becomes increasingly difficult since established cooling methods are limited in reaching even lower temperatures while preserving high densities \cite{timmermans_degenerate_2001}. 
%{\bf hier brauchen wir referenzen, z.B. Phys. Rev. Lett. 87, 240403 und am besten noch ein paar experimentelle}. 
This has been a major obstacle on the route to exploring novel phases of matter. 

Over the past decades, several techniques to reach even lower temperature have been proposed. This includes, for example, entropy reduction by separating high and low entropy regions in the trap \cite{popp_ground-state_2006,capogrosso-sansone_monte_2008,bernier_cooling_2009}, by introducing disorder  \cite{unal_cooling_2017}, utilizing heat reservoirs in form of a different state or species \cite{ho_squeezing_2009}, or by inter-layer entropy exchange in bilayer systems \cite{kantian_dynamical_2018}.
Novel cooling techniques have been experimentally demonstrated, for example, in the Hubbard model by isolating a low entropy region via potential shaping \cite{chiu_quantum_2018}, and by creating site-alternating heat reservoirs via the introduction of a superlattice in one spatial direction \cite{yang_cooling_2020}.
%{\bf: ergänzen für Bosonen: Phys. Rev. A 74, 013622 (2006)., B. Capogrosso-Sansone et al., Phys. Rev. A 77, 015602. Hier gibt es noch eine Liste weiterer Paper mit Kühl-Proposals:
%https://journals.aps.org/prl/abstract/10.1103/PhysRevLett.120.060401, 
%https://journals.aps.org/prl/abstract/10.1103/PhysRevLett.120.243201, 
%https://journals.aps.org/pra/abstract/10.1103/PhysRevA.98.033607, 
%https://iopscience.iop.org/article/10.1088/1367-2630/aa5e7b, 
%https://journals.aps.org/prl/abstract/10.1103/PhysRevLett.115.243002 -- kannst Du mal schauen, was davon noch zitierwürdig wäre? Es wäre schon sinnvoll auf verschiedene Techniken zu verweise, damit (1) der Gutachter versteht, dass dies ein wichtiges Feld ist und (2) wir nicht nur gegen eine technik gemessen werden. Es wäre auch gut nochmal nach weiteren experimentellen Arbeiten dazu zu suchen.}
%

The  concept of evaporative cooling is to remove or separate particles with an energy higher than the average energy per particle from the sample.  This leads to a reduction of both temperature and atoms, however, when performed efficiently, it will increase the  phase-space density and therefore lower the entropy of the gas. This paves the way towards the desired progression towards low-entropy states. In magnetic traps, evaporation is achieved by radio-frequency induced transitions to anti-trapped states and in optical traps by lowering the potential depth. Both methods have been successfully implemented, however, both also have limitations. Evaporative cooling in magnetic traps can be challenging for mixtures of atoms in different spin states, which is a common problem when cooling Fermi gases, and evaporative cooling in optical traps faces the challenge of reducing confinement at the same time as lowering the trap depth. This poses problems for maintaining high densities and staying in the runaway regime of evaporative cooling.
%, cooling in optical dipole traps has enabled the efficient cooling of various state mixtures while freely applying additional magnetic fields to further control to sample. Evaporation cooling in an optical dipole trap works by ramping down the trap depth, which favorably removes atoms from the high energy tail resulting in a decrease in temperature \cite{ohara_scaling_2001, ohara_observation_2002, granade_all-optical_2002}. As both the lowered trap depth as well as the removal of atoms results in a decrease in density, it requires a lot of optimization to stay in the efficient runaway regime \cite{granade_all-optical_2002}. Many ideas have been demonstrated to increase the efficiency by maintaining high density through manipulating the trap geometry via gradients \cite{hung_accelerating_2008, clement_all-optical_2009}, altered trap sizes \cite{kinoshita_all-optical_2005} or crossed traps \cite{arnold_all-optical_2011}.

\begin{figure}
	\includegraphics[width=\columnwidth]{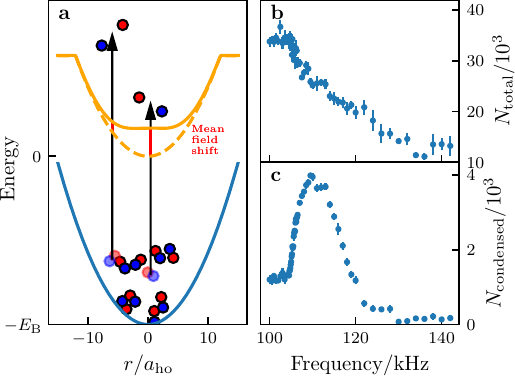}	
	\caption{\textbf{a} Composite dimers with binding energy $E_{\mathrm{B}}$ are confined in a harmonic trapping potential (blue line) and constitute a partly condensed quantum gas. A periodic modulation of their interaction strength dissociates  them into pairs of fermions with an energy dependent on the modulation frequency and the mean-field energy shift. The latter is proportional to the density of remaining dimers and fermions and introduces a position-dependent dissociation threshold. \textbf{b}  Varying the modulation frequency leads to evaporation of atoms if the modulation frequency is significantly above the dissociation threshold. \textbf{c} If the evaporation is selectively for the dimers with high energies, the number of atoms in the condensate increases contrary to the total decrease of atom number. If the evaporation is non-specific, i.e. far above the dissociation threshold, the condensate is diminished along with the general loss of particles.}
	%\caption{Density dependent dissociation frequency for composite dimers. The upper plot shows a sample of composite dimers confined in an harmonic trap (blue line) which are selectively dissociated with a spatially varying dissociation frequency to the unbound state (orange line). To dissociate a pair at a specific location, the bound-state energy $E_\mathrm{B}$ plus a density dependent term $E_\mathrm{D}-E_\mathrm{B}$ (red vertical segment) have to be applied.}
	\label{Fig1}
\end{figure}

In this work, we demonstrate a novel cooling method for a strongly-interacting quantum gas on the repulsive side of a magnetically-controlled Feshbach resonance. At low temperature the gas comprises of dimers with a binding energy $E_{\mathrm{B}}$. We apply a time-periodic modulation of the interaction strength with a frequency near to or above the binding energy in order to dissociate dimers with a certain energy and selectively remove them from the sample. The matrix element for the transition depends on the binding energy, the density of states and, last but not least, on the interaction in both initial and final states. We utilize the spatial dependence of the mean-field interaction in a harmonic trapping potential in order dissociate molecules spatially selectively. This allows us to discriminatively  remove particles with an energy higher than $k_{\mathrm{B}}T$ from the sample. We demonstrate that this method leads to cooling and we show an increase of the condensate fraction of a partially condensed system. Importantly, unlike radio-frequency induced evaporation, our method does not involve a third spin state, thereby eliminating three-body losses, and unlike evaporation in optical dipole traps, it does not soften the trapping potential and slow down thermalization.

Experimentally, we prepare a quantum gas in the lowest and third-lowest hyperfine states $\ket{1}$ and $\ket{3}$ of $^{6}\textrm{Li}$ in a crossed-beam optical dipole trap \cite{behrle_higgs_2018} with trap frequencies $\left(\omega_x, \omega_y, \omega_z\right) = 2\pi \times \left(102, 144, 232\right)\,\mathrm{Hz}$. We obtain atom numbers of $N_{\sigma} \sim 5\cdot10^{4}$ -- $1.5\cdot10^{5}$ per spin state $\sigma$ at a temperature of $T / T_{\mathrm{F}} \sim 0.1$.
The interaction strength and binding energy are adjusted by a magnetically-controlled Feshbach resonance located at $690.43\,\mathrm{G}$ with a width of $122.3\,\mathrm{G}$ \cite{zurn_precise_2013}. 
We calibrate the magnetic field using the  $\ket{1}$ to $\ket{2}$ hyperfine transition in a spin-polarised Fermi gas, and the binding energy of the dimer state energy is determined from the magnetic field and the interatomic interaction potential \cite{gao_binding_2004,gribakin_calculation_1993, mitroy_semiempirical_2003}.
In order to induce pair-breaking excitations, we apply a periodically modulated magnetic field with an amplitude of up to $A_{\mathrm{mod}} \sim 2\,\mathrm{G}$ and a frequency of up to $\nu_{\mathrm{mod}} \sim 300.0\,\mathrm{kHz}$ using a specially designed solenoid \cite{kell_compact_2021}. In contrast to a dissociation scheme by radiofrequency modulation \cite{gupta_radio-frequency_2003}, the dissociation by magnetic field modulation does not involve transfer to a third hyperfine state and, moreover, does not excite free atoms. After an excitation of usually $300\,\mathrm{ms}$, we introduce a waiting time of $200\,\mathrm{ms}$ in order to ensure thermalisation of the sample. During thermalisation, there is a small particle loss since the decay time constant of the condensate is $1\,\mathrm{s}$.
Detection is performed by absorption imaging either in-situ or after rapid-ramp projection \cite{regal_observation_2004} in  time-of-flight.

In Fig.~1a, we show the principle of the cooling scheme. The initial state is a gas predominantly containing dimers in a harmonic potential.  Between two dimers, the s-wave scattering length is $a_{\mathrm{dd}}=0.6\, a$  \cite{petrov_weakly_2004}, where $a$ is the inter-atomic scattering length and this gives rise to the mean-field potential for the dimers $\propto a_{\mathrm{dd}}n$. A time-periodic modulation of the magnetic field at a frequency near the binding energy breaks the dimers into a pair of free atoms. The energy threshold for the dissociation is determined by the dimer binding energy and the  difference between initial-state and final-state interactions. The final-state interaction is determined by the scattering length between the dissociated atoms and the remaining dimers, which is $a_{\mathrm{ad}}=1.18\,a$ between a dimer and an atom \cite{skorniakov_three_1957}. For modulation frequencies above the dissociation threshold, additional kinetic energy is given to the dissociation fragments, which may evaporate from the trap. The key idea of the cooling scheme is to exploit the change of the mean-field energy before and after dissociation of a dimer into atoms, which is
$\delta E = (3a_{\mathrm{ad}}-a_{\mathrm{dd}}) 2 \pi \hbar^2 n_\sigma / m_{\mathrm{Li}}$ \cite{combescot_normal_2007,giorgini_theory_2008,pitaevskii_bose-einstein_2016}. As one can see from Fig.~1b, the number of removed particles is a monotonous function of the modulation frequency. The dissociation is position sensitive due to the mean-field shift. For low density -- which equals high energy of dimers in the trap -- the dissociation threshold is close to the binding energy $E_B$. In contrast, for high densities, the dissociation threshold is shifted due to the mean-field interaction to larger values and more particles are removed from the trap. By explicitly dissociating dimers in the high-energy tails of their Boltzmann distribution, which corresponds to low densities in a harmonic trap,  we are  able to effectively reduce the energy per particle, cool the sample, and enhance the phase-space density, see Figure 1c.

\begin{figure}
	\includegraphics[width=\columnwidth]{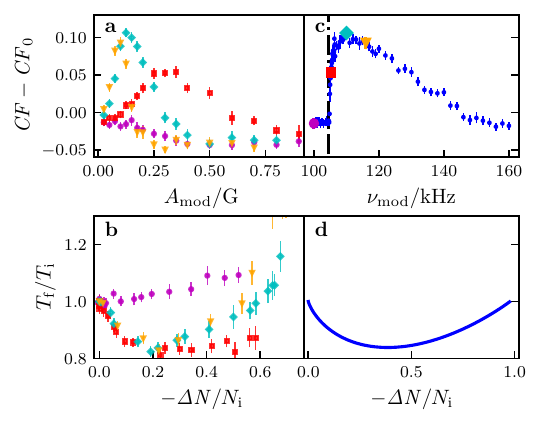}
	\caption{ \textbf{a} Change of the condensate ($\mathit{CF}$) fraction as a function of the modulation amplitude $A_{\mathrm{mod}}$ for four different modulation frequencies $\nu_{\mathrm{mod}}$ of $100.0\,\mathrm{kHz}$ (purple, circle), $105.2\,\mathrm{kHz}$ (red, square), $110.0\,\mathrm{kHz}$ (cyan, diamond) and $116.0\,\mathrm{kHz}$ (orange, triangle). The sample has a binding energy of $104.5(5)\,\mathrm{kHz}$. \textbf{b} Variation of the temperature with the fractional atom loss $-\mathit{\Delta N}/N_{\mathrm{i}}$ for the same modulation amplitudes and frequencies as in \textbf{a}. \textbf{c} Maximum relative condensate fraction for each driving frequency. The four frequencies from \textbf{a} are highlighted in the corresponding colors. The dashed-dotted line shows the binding energy. The error bars show the standard error.  \textbf{d} Theoretical estimate for the cooling as a function of atom number loss using an interacting Bose gas model, see main text.}
	\label{Fig2}
	%% Frequencies are 104.5e3, 105.1e3, 108e3, 112e3
\end{figure}

In Fig.~2a we show the change of condensate fraction due to the periodic modulation of the magnetic field with respect to modulation strength and modulation frequency at a magnetic field of $640.31(6)\,\mathrm{G}$, corresponding to a binding energy of $104.5(5)\,\mathrm{kHz}$. We observe distinct regimes of cooling and heating depending on frequency and modulation amplitude. For a modulation at or below the frequency of the binding energy we observe no increase in condensate fraction, but instead the condensate decays with increasing modulation strength. In Fig.~2b we show the corresponding change of the temperature of the gas, which, again, shows the regimes of cooling and heating as above. For a more meaningful comparison later, the modulation amplitude axis is replaced by the corresponding atom loss $-\mathit{\Delta N}/N_{\mathrm{i}}$, with respect to the initial atom number $N_{\mathrm{i}}$. For each modulation frequency we determine the maximum achieved condensate fraction and plot an overview of all maxima in Fig.~2c. We notice that cooling is possible over a wide range of modulation frequencies above the energy corresponding to the binding energy. %First, we notice an asymmetric profile with sharp onset at the bound-state energy, which is typical for dissociation processes into a continuum of states \cite{schunck_determination_2008, regal_creation_2003, regal_probing_2005, bartenstein_precise_2005}. The exact lineshape depends on the trap geometry and the overlap between the initial molecular wavefunction with the wavefunction of the resulting atom pair \cite{regal_creation_2003}. 

We also have investigated how rapid the cooling method works. The data of Figure 2 are for a modulation time of 300\,ms for which we can systematically study a broad range of modulation amplitudes. If we instead focus on the strongest possible modulation amplitude, we can significantly shorten the cooling time. To this end, we have determined that the modulation time to achieve maximum cooling scales as $t_{\mathrm{mod}} \propto 1 / \sqrt{A_{\mathrm{mod}}}$ and we can achieve the same peak increase of the condensate fraction as in Figure 2 with a cooling time of $t_{\mathrm{mod}}=9.3\,\mathrm{ms}$, followed by a $1/e$ thermalisation time of $\sim 20\,\mathrm{ms}$.

In order to explain our cooling scheme and the dependence of the cooling effect on the modulation frequency, we have developed a simplified model based on the step model of evaporative cooling \cite{ketterle_evaporative_1996}. We describe the initial density distribution of the dimers by a trapped interacting Bose gas. To this end, we employ our independently measured atom number, temperature (see below), and trap frequencies, and we compute the internal energy for the initial state. As the cooling step, we remove a certain fraction (depending on the modulation frequency) of dimers from the sample, and, finally, we compute the internal energy and temperature of the remaining dimers. For the comparison between the model and our data, we focus on the fractional atom loss $-\mathit{\Delta N}/N_{\mathrm{i}}$ and show that we can reach a 15\% reduction of the initial temperature for a fractional atom loss of 30\%. The magnitude of the cooling effect is in agreement with the experimental observations in Fig.\ 2b. In the experimental data, we actually also observe heating for certain parameter regimes. This effect is not captured by our model and could be caused by several reasons: Firstly, not all dissociated dimers will leave the trapping region but some some could remain with excess kinetic energy from above-threshold dissociation. Secondly, our model ignores the dimer condensate and dissociation from the dimer condensate removes over-proportionally many dimers with zero kinetic energy.

%We compare our results with a model of evaporative cooling. We start from the measured (?) in-situ density distribution and compute the number of dissociated dimers and their average energy  for a given modulation frequency and modulation amplitude. From this, we compute the change of temperature of the remaining sample. Obviously, the change of temperature is a non-monotonic function of the frequency: At the binding energy, very few dimers with very high potential energy are dissociated since the densities for which the mean-field shift is near-zero are very small. As we increase the modulation frequency above the binding energy, we dissociate dimers more and more towards the center of the trap. This increases the total number of dimers dissociated and also reduces the average energy of the dissociated molecules. Initially, this leads to a cooling effect, however, if we dissociate too close to the trap center, the large loss of very cold dimers eventually terminates the cooling effect. In Figure 2c(?) we show our simulation of the cooling using an interacting  Bose gas (?) {\bf can we do this with our actual measured density distribution? This would be better.} and find a20\% reduction of the temperature. In Figure 2d (?), we show the corresponding experimental data, which confirm the expected reduction of temperature for this parameter set. {\bf hier kannst Du gerne noch mehr Details zu der SImulation einfügen}

\begin{figure}
	\includegraphics[width=\columnwidth]{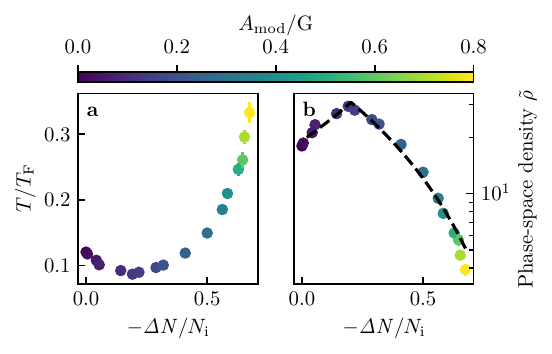}
	\caption{Cooling of a sample with binding energy $104.5(5)\,\mathrm{kHz}$ at a modulation frequency of $110.0\,\mathrm{kHz}$. The colorbar highlights the applied modulation amplitude. \textbf{a} Shows the atom number with respect to the reduced temperature after cooling. \textbf{b} Phase-space density vs.\ the atom number in a logarithmic plot. The dashed line displays the piece-wise exponential fit to determine the cooling efficiency. Errorbars are standard errors and might be hidden by datapoints.}
	\label{Fig3}
\end{figure}

In order to quantify the cooling efficiency, we additionally measure the phase-space density versus atom number at the modulation parameters corresponding to maximum enhancement of the condensate fraction. To this end, we take high-intensity absorption images \cite{hueck_calibrating_2017} of the sample to reconstruct the 3D density profile $n_{\sigma}(r)$ per spin state via the inverse Abel transformation \cite{shin_observation_2006}. We use the density at the trap center $n \equiv n_{\sigma}(r=0)$ for the phase-space-density $\tilde{\rho}=n \lambda_{\mathrm{dB}}^3$, with $\lambda_{\mathrm{dB}}$ being the thermal de-Broglie wavelength.
%Equally, the center density is used to compute the Fermi energy $E_{\mathrm{F}} = \hbar^2 k_{\mathrm{F}}^2 / \left( 2 m \right)$ and momentum $k_{\mathrm{F}} = \left(6 \pi^2 n_{\sigma}\right)^{1/3}$ relevant for the dimensionless interaction parameter $1 / \left( k_{\mathrm{F}} a\right)$. {\bf wofür wird 1/kFa gebraucht? \textcolor{red}{Um die Methode in der ueblichen sprache des BEC BCS crossovers einzuordnen. Wir geben den Wert auch zweimal an.} ok wir lassen es jetzt mal so dring, aber ich bin nocht überzegt, dass wir es wirklich brauchen.} 
In order to determine the temperature, we fit the outer wings of the reconstructed density profile to the virial expansion of the equation of state \cite{link_machine_2023}. %At these magnetic fields we take measurements at several modulation amplitudes $A_{\mathrm{mod}}$.
In Figure 3a we show the reduced temperature $T/T_{\mathrm{F}}$ and in Figure 3b the phase-space-density $\tilde{\rho}$ with respect to the fractional atom loss $-\mathit{\Delta N}/N_{\mathrm{i}}$ at an exemplary magnetic field of $640.31(6)\,\mathrm{G}$ (the same as in Figure 2c) and a modulation frequency of $110.0\,\mathrm{kHz}$. For small  modulation amplitudes we observe a reduction of temperature and an increase of phase-space density and for high modulation amplitude, the fractional atom loss overwhelms the cooling effect leading to heating.

\begin{figure}
	\includegraphics[width=\columnwidth]{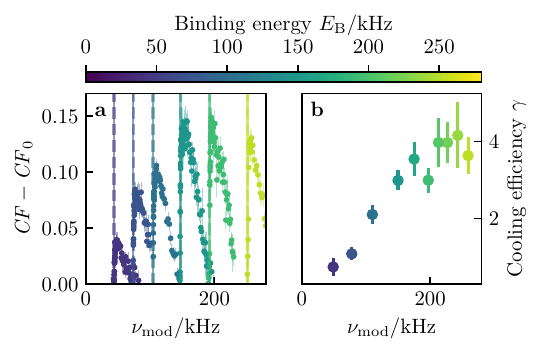}
	\caption{Cooling of samples from the bosonic side of the BEC/BCS crossover towards unitarity. The position in the crossover is given by the binding  energy $E_{\mathrm{B}}$ in the color map. \textbf{a} Maximum gain of the condensate fraction as determined in Fig.~2b. The vertical dashed lines show the binding energy for the chosen magnetic field. \textbf{b} Cooling efficiency as determined in Fig.~3b.}
	\label{Fig4}
\end{figure}

The efficiency of evaporative cooling is measured by the parameter $\gamma = -\log(\tilde{\rho}_{\mathrm{f}}/\tilde{\rho}_{\mathrm{i}}) / \log(N_{\sigma, \mathrm{f}} / N_{\sigma, \mathrm{i}})$ \cite{ketterle_evaporative_1996}, which is a metric of how much the phase-space density increases $\tilde{\rho}_{\mathrm{i}} \rightarrow \tilde{\rho}_{\mathrm{f}}$ while reducing the atom number $N_{\sigma, \mathrm{i}} \rightarrow N_{\sigma, \mathrm{f}}$. To deduce the cooling efficiency $\gamma$ we fit a piece-wise exponential function with the efficiency as the exponent to the data, which can be seen as curves in the logarithmic plot in Figure 3b. 
A positive slope indicates the regime of efficient cooling and a negative slope a regime that should be avoided for evaporation. The maximum efficiency for a given binding energy depends on both modulation frequency and amplitude. In the samples with large binding energy, say $E_{\mathrm{B}}\sim 200\,\mathrm{kHz}$, we reach an efficiency of $ \gamma \sim 4$, which matches or exceeds the best reported values  of $\gamma$ in magnetic and optical traps \cite{ketterle_evaporative_1996, granade_all-optical_2002,olson_optimizing_2013, roy_rapid_2016, clement_all-optical_2009}.
%A positive slope denotes a negative cooling efficiency, i.e. a regime that should be avoided for evaporation,  and a negative slope is the regime in which the cooling is efficient.  The maximum efficiency for a given binding energy depends on both modulation frequency and amplitude. In the samples with large binding energy, say $E_{\mathrm{B}}\sim 200\,\mathrm{kHz}$, we reach an efficiency of $ \gamma \sim 4$, which matches the best reported values from other techniques including bosonic and fermionic systems, such as evaporation in magnetic and optical-dipole traps \cite{ketterle_evaporative_1996, granade_all-optical_2002}, as well as evaporation in optical-dipole traps which are modified and optimized to stay closer to the efficient runaway regime \cite{olson_optimizing_2013, roy_rapid_2016, clement_all-optical_2009}.

In Figure 4, we show how the cooling works as we vary the binding energy all the way from the bosonic side of the BEC/BCS crossover to unitarity. Generally, we observe that cooling works best for deeply-bound dimers and the efficiency reduces when approaching the  unitarity regime. Figure 4a shows the maximum change of the condensate fraction as determined in Fig.~2c, and Figure 4b shows the maximum cooling efficiency. For the lowest magnetic fields, i.e. the largest binding energy, we observe an increase of the condensate fraction by $\sim 0.15$ and a cooling efficiency of $\sim 4$. Getting closer to the unitary regime  the cooling  becomes less and eventually, at an interaction parameter of $1/(k_{\mathrm{F}}a)  \sim 0.7$, the condensate fraction does not increase and the cooling efficiency drops below unity. There are at least two effects, which contribute to this behaviour: Firstly, a small dimer binding energy makes the excitation less spatially selective, and, secondly, with the increasing dimer size towards the unitary regime many-body correlations play an increasingly important role, which changes the  spectral line shape of the excitation \cite{schunck_determination_2008}.

% Papers showing of different efficiencies
% All Optical 2002 - All-Optical Production of a Degenerate Fermi Gas S. R. Granade DOI: 10.1103/PhysRevLett.88.120405, around 3
% Evaporative Cooling of Trapped Atoms 1996 - Ketterle and Druten - Nice Table at the end - Efficiencies between 0.8 - 3.0

%Interaction Parameters including stderr fit from density fit.
%637.0 2.568103815706763 0.0142507805395524
%638.0 2.46618591152724 0.011159279378237
%639.0 2.37759218127974 0.0116034263728885
%640.0 2.2040120734795643 0.0113638356371409
%641.0 2.102846050808429 0.0088372772562082
%643.0 1.837274543209728 0.0081752575753213
%645.0 1.7127784072061811 0.0076281247210692
%650.0 1.2020417388224731 0.0093215256039588
%655.0 1.003216494693345 0.0040425508708076
%662.0 0.6971929757214922 0.0022411269307288

While the results of the manuscript have been shown for a gravity tilted trap in a $\ket{1}$-$\ket{3}$ mixture, we  observe  cooling  in a gravity-compensated trap and for a $\ket{1}$-$\ket{2}$ mixture with comparable efficiency. Hence, the cooling method should be ready to implement in a large variety of systems.

%\begin{figure}
%	%\includegraphics[width=\columnwidth]{figures/Freq vs Cooling - manuscript.pdf}
%	\includegraphics[width=\columnwidth]{Freq vs Cooling - manuscript - no colorbar.pdf}
%	\caption{Maximum cooling efficiencies across the BEC regime. For each bound state energy the cooling efficiency at the frequency of the maximum condensate fraction in Fig.~\ref{Fig2} b) is shown. The errorbars are the standard errors of the piece-wise fits shown in Fig.~\ref{Fig4}. } 
%	\label{Fig5}
%\end{figure}

In conclusion, we demonstrate a novel cooling scheme for a gas in the BEC/BCS crossover regime based on the targeted removal of high energy atoms from an optical trap by dimer dissociation.  Samples can be cooled without changing the trap geometry which simplifies staying in the run-away regime and to reach very high efficiencies.

This work has been supported by the Deutsche Forschungsgemeinschaft through SFB/TR 185 (project B4) and the Cluster of Excellence Matter and Light for Quantum Computing (ML4Q) EXC 2004/1 – 390534769. 

%\section*{APPENDIX}

%\subsection*{Something Maybe}
%Potential Appendix Material here


\end{document}